\documentclass[prb,onecolumn,nobibnotes,groupedaddress, natbib]{revtex4}

\usepackage{enumerate}
\usepackage{amssymb}
\usepackage{amsmath}
\usepackage{mathtools}
\usepackage{amssymb}
\usepackage{braket}
\usepackage[hidelinks]{hyperref}
\usepackage{pdfpages}
\usepackage{epstopdf}

\DeclareMathOperator*{\argminA}{arg\,min}

\begin{document}

\title{A new ``gold standard'': perturbative triples corrections in unitary coupled cluster theory and prospects for quantum computing}

\thanks{This manuscript has been authored by UT-Battelle, LLC, under Contract No.~DE-AC0500OR22725 with the U.S.~Department of Energy. The United States Government retains and the publisher, by accepting the article for publication, acknowledges that the United States Government retains a non-exclusive, paid-up, irrevocable, world-wide license to publish or reproduce the published form of this manuscript, or allow others to do so, for the United States Government purposes. The Department of Energy will provide public access to these results of federally sponsored research in accordance with the DOE Public Access Plan.}
%\date{}
\author{Zachary W. Windom$^{1, 2}$, Daniel Claudino$^2$\footnote{\href{mailto:claudinodc@ornl.gov}{claudinodc@ornl.gov}}, and Rodney J. Bartlett$^1$}
\affiliation{$^1$Quantum Theory Project, University of Florida, Gainesville, FL, 32611, USA \\
$^2$Computational Sciences and Engineering Division,\ Oak\ Ridge\ National\ Laboratory,\ Oak\ Ridge,\ TN,\ 37831,\ USA}

\begin{abstract}
  A major difficulty in quantum simulation is the adequate treatment of a large collection of entangled particles, synonymous with electron correlation in electronic structure theory, with coupled cluster (CC) theory being the leading framework in dealing with this problem. Augmenting computationally affordable low-rank approximations in CC theory with a perturbative account of higher-rank excitations is a tractable and effective way of accounting for the missing electron correlation in those approximations. This is perhaps best exemplified by the “gold standard” CCSD(T) method, which bolsters the baseline CCSD with effects of triple excitations using considerations from many-body perturbation theory (MBPT). Despite this established success, such a synergy between MBPT and the unitary analog of CC theory (UCC) has not been explored. In this work, we propose a similar approach wherein converged UCCSD amplitudes, which can be obtained on a quantum computer, are leveraged by a classical computer to evaluate energy corrections associated with triple excitations - leading to the UCCSD[T] and UCCSD(T$^*$) methods. The rationale behind these choices is shown to be rigorous by studying the properties of finite-order UCC energy functionals. Although our efforts do not support the addition of the fifth-order contribution as in the (T) correction, comparisons are nevertheless made using a hybrid UCCSD(T) approach. We assess the performance of these approaches on a collection of small molecules, and demonstrate the benefits of harnessing the inherent synergy between MBPT and UCC theories. 
\end{abstract}

%% requires nobibnotes 

\maketitle

%\section{Introduction}
The exact solution of the time-independent, non-relativistic Schr\"{o}dinger equation is the ``holy grail" of quantum chemistry, as \textit{ab initio} prediction of several important molecular and materials properties becomes immediately accessible.\cite{bartlett1989coupled} Unfortunately, full configuration interaction (FCI), i.e., the accounting of all possible electronic configurations in a one-particle basis, scales combinatorially with system size, meaning that the exact solution is beyond reach for the vast majority of chemical space. Nevertheless, methods based on low-rank coupled-cluster (CC) theory have the advantage of converging rapidly to the FCI limit in polynomial time,\cite{bartlett2007coupled} and therefore are of immense value to the computational chemistry and materials science communities.\cite{zhang2019coupled} In fact, systematic convergence to FCI is assured by considering higher-rank cluster operators into the ans\"{a}tze albeit at increasing computational cost.\cite{shavitt2009many}

Contrary to approaches centered around expectation values, CC lends itself to a series of residual equations emerging from projections of the Schr\"{o}dinger equation onto the space of excitations out of the reference function, typically Hartree-Fock (HF), but applicable to any single determinant that overlaps with the exact wavefunction. One way to circumnavigate the mounting intractability in including arbitrarily high-orders of the cluster operator is to introduce corrections \textit{post hoc} based on some flavor of perturbation theory. The most prominent of these methods is the perturbative energy correction for triple excitations with infinite-order single (S) and double (D) excitations that results in the CCSD(T) method\cite{urban1985towards,raghavachari1989fifth,watts1993coupled} - the so-called ``gold standard" of quantum chemistry. And, on the same token, a similar philosophy can be used to bolster - for example, CCSDT - by incorporating a perturbative treatment of missed quadruple excitations.\cite{kucharski1989coupled} This framework ultimately culminates in a hierarchy of methods whose focus is to provide a perturbative estimate of electron correlation associated with higher-rank cluster operators that are explicitly omitted once the cluster operator has been truncated.\cite{bartlett1990non} Methods based on the factorization theorem of MBPT have also been proposed and investigated,\cite{kucharski1986fifth} showing that further reduction in calculation cost can be achieved while simultaneously providing some estimate of higher-order correlation effects.\cite{kucharski1998efficient, kucharski1998noniterative,kucharski1998sixth}

A simplistic view of the CC ansatz would define it as an exponential map of excitation operators, which is used in its predominant projective variant. This ``simplicity" arises by virtue of the natural truncation of the Baker-Campbell-Hausdorff expansion of the similarity-transformed Hamiltonian, and it comes at the expense of a loss of unitarity/variationality. However, other perspectives have investigated alternative CC ans\"{a}tze\cite{szalay1995alternative} in the pursuit of satisfying exact conditions,\cite{bartlett1989some,szalay1995alternative} such as the generalized Hellman-Feynman (GHF) theorem.\cite{helgaker1984second} The original expectation value formalism\cite{bartlett1988expectation} - which, in the limit, converges toward variational CC\cite{van2000benchmark,cooper2010benchmark} - and the unitary CC (UCC) ans\"{a}tze\cite{bartlett1989alternative} fall into this category. By design, these ans\"{a}tze are arguably more suitable for the calculation of properties as compared to the standard CC formulation. Unfortunately, they inherently scale as FCI since the Hamiltonian-cluster commutator expansion does not truncate.\cite{shavitt2009many} Again, perturbation theory can be used to ``pick" a suitable truncation point for tractability,\cite{taube2006new} although admittedly methods that are truncated at low-orders do not necessarily provide results that are comparable to similar, standard CC counterparts.\cite{watts1989unitary,byrd2014correlation} Alternatively, truncations based on commutator-rank have also been explored.\cite{liu2021unitary,liu2022quadratic}

This impasse between projective and alternative ans\"{a}tze may potentially be solved with the emergence of quantum computing paradigms, which would enable UCC without resorting to arbitrary truncation. This is because the UCC ans\"{a}tze can be effectively encoded as a series of gate-based operations acting on an \textit{easy-to-prepare} state,\cite{anand2022quantum,chen2021quantum} e.g. HF. In fact, there is supporting evidence that untruncated UCC theory provides more accurate results than the equivalent, standard CC method.\cite{cooper2010benchmark} Nevertheless, fundamental issues inhibit routine UCC calculations on a quantum computer; notably, circuit width and depth – either of which ultimately restrict the maximal rank of the cluster operator.\cite{claudino2022basics} This is clearly unfortunate since it is known that the only way CC/UCC converges toward the exact solution of the Schr\"{o}dinger equation (in a basis set) is by adding the abovementioned higher-rank cluster operators. A further point of consideration lies in the distinction between the complete, non-terminating UCC ansatz and the Trotterized, or so-called \textit{disentangled}, counterpart, with the latter rising from known product formulas and proven exact under certain conditions.\cite{evangelista2019exact}

To address the lack of a framework enabling perturbative corrections for the UCC ansatz, we explore potential synergies between MBPT and UCC theory which have not been scrutinized until now. By studying the properties of finite-order UCC equations, we propose a pathway towards perturbative corrections to the infinite-order UCCSD energy designed to recover the missed effects of higher-rank excitations, which we illustrate by introducing the UCCSD[T] and UCCSD(T$^*$) methods, unitary analogs of the pioneering perturbative accounting for triple excitations.\cite{urban1985towards} We show that such an approach is a robust way of recovering electron correlation that is missed by restricting the unitary cluster operator. By employing what can be seen as a post-processing step born from rigorous theory, our results reinforce the perspective that resource-efficient interplay between quantum and classical computers should be harnessed to achieve more accurate results without imposing extra burden on current, fragile quantum computers.

We motivate the problem using semi-canonicalized orbitals for a general reference function, e.g., non-HF. The normal-ordered Hamiltonian is then of the form
\begin{equation}
    H_N=\underbrace{\bigg(\sum_{p}\epsilon_{pp}\{p^{\dagger}p\}\bigg)}_{\text{$f_N$}} + \underbrace{\bigg(\sum_{ia}f_{ia}\{i^{\dagger}a\}+\sum_{ai}f_{ai}\{a^{\dagger}i\} +\frac{1}{4}\sum_{pqrs}\braket{pq||rs}\{p^{\dagger}q^{\dagger}sr\}\bigg)}_{\text{$W_N$}},
\end{equation}
where indices $a,b,c, d\cdots$, $i,j, k, l\cdots$, and $p,q,r,s\cdots$ specify virtual, occupied, and arbitrary spin-orbitals. Note that  the perturbation $W_N$ now contains the occupied/virtual blocks of the Fock operator.

Standard CC theory begins by defining the form of the cluster operator, $T$, 
\begin{equation}
    T=T_1+T_2+T_3+\cdots,
\end{equation} where each $T_n$ can be expressed in the language of second quantization as
\begin{equation}
    T_n=\frac{1}{(n!)^2}\sum_{\substack{a, b, c, \cdots \\ i, j, k, \cdots}} t_{ijk\cdots}^{abc\cdots}\{a^{\dag}i b^{\dag}jc^{\dag}k \cdots \}.
\end{equation}
Once an appropriate level of cluster restriction has been chosen, the unitary cluster operator can be defined as
\begin{equation}\label{eq:tauDef}
    \tau=T-T^{\dag},
\end{equation} where our working assumption is that we are using real orbitals, hence $t_{ij\cdots}^{ab\cdots*}=t_{ij\cdots}^{ab\cdots}$. In this context, the Schr\"{o}dinger equation becomes 
\begin{equation}
\label{eq:tauEqns}
    H_Ne^{\tau}\ket{0}= \big( H_Ne^{\tau}\big)_C\ket{0} = \Delta E e^{\tau}\ket{0},
\end{equation} 
%where the equivalence 
%\begin{equation}\label{eq:tauEqns}
%    \big( e^{\tau^{\dag}}H_N e^{\tau}\big)\ket{0} = \big( e^{-\tau}H_N e^{\tau}\big)\ket{0} =\big( H_Ne^{\tau}\big)_C\ket{0}=\Delta E \ket{0}
%\end{equation} is noted, 
with $C$ indicating a restriction to connected diagrams and $\Delta E = E_\text{CC} - E_\text{HF}$ being the correlation energy. We point out that if we follow the traditional CC route of projecting Equation \ref{eq:tauEqns} onto elements of the excitation manifold, the resulting residual equations will not terminate. Therefore, we have to pick a point to truncate the resulting expressions based on some specified criteria. Unlike prior work on the topic,\cite{bartlett1989alternative, Kutzelnigg1991} we define orders in terms of $W_N$ assuming a non-canonical HF reference, denoted by $\ket{0}$. In other words, $f_N$ is zeroth-order and $W_N$ arises in first-order of MBPT and therefore both $\tau_1$ and $\tau_2$ show up at first-order whereas the remaining higher-order operators, $\tau_n$, arise in the $(n-1)$-order wavefunction of MBPT.  

For the purposes of this work, our starting point is the complete, fourth-order UCC(4) energy functional

\begin{equation}\label{eq:e4functional}
\begin{split}
    %\Delta E^{[4]} 
    \Delta E(4)=& -\braket{0|\frac{1}{3!}\bigg((T_1^{\dagger})^2T_1f_NT_1+\text{h.c.}+(T_2^{\dagger})^2T_2f_NT_2+\text{h.c.}\bigg)+\frac{1}{2}\bigg(T_3^{\dagger}T_1f_NT_2+\text{h.c.}+T_3^{\dagger}T_2f_NT_1+\text{h.c.}\bigg)|0}\\
        &-\frac{1}{3}\braket{0|T_1^{\dagger}T_2^{\dagger}T_1f_NT_2+\text{h.c.}+T_2^{\dagger}T_1^{\dagger}T_2f_NT_1+\text{h.c.}|0}\\
        &-\braket{0|\frac{1}{3!}\Big((T_1^{\dagger})^2T_1W_N+\text{h.c.}\Big) +\frac{1}{3!}\Big((T_2^{\dagger})^2T_2W_N+\text{h.c.}\Big)+\frac{1}{2}\Big(T_3^{\dagger}T_1W_N+\text{h.c.}+T_3^{\dagger}T_2W_N+\text{h.c.}\Big)|0} \\
        &-\frac{1}{3}\braket{0|T_1^{\dagger}T_2^{\dagger}T_1W_N+\text{h.c.}+T_2^{\dagger}T_1^{\dagger}T_2W_N+\text{h.c.}|0}-\frac{1}{6}\braket{0|T_2^{\dagger}T_1T_1^{\dagger}f_NT_2+\text{h.c.}+T_2^{\dagger}T_1T_1^{\dagger}W_N+\text{h.c.}|0}\\
        &+\frac{1}{4}\braket{0|(T_1^{\dagger})^2f_NT_1^2+(T_2^{\dagger})^2f_NT_2^2|0}+\braket{0|T_3^{\dagger}f_NT_3|0}+\braket{0|T_3^{\dagger}f_NT_1T_2+\text{h.c.}|0}+\braket{0|T_1^{\dagger}T_2^{\dagger}f_NT_1T_2|0}\\
        &+\frac{1}{2}\braket{0|(T_1^{\dagger})^2W_NT_1+\text{h.c.}+(T_2^{\dagger})^2W_NT_2+\text{h.c.}|0}+\braket{0|T_3^{\dagger}W_N(T_1+T_2)+\text{h.c.}|0}+\braket{0|T_2^{\dagger}T_1^{\dagger}W_N(T_1+T_2)+\text{h.c.}|0}\\
        &-\frac{1}{2}\braket{0|T_2^{\dagger}T_1W_N(T_1+T_2)+\text{h.c.}|0}+\frac{1}{2}\braket{0|(T_1^{\dagger})^2W_NT_2+\text{h.c.}|0}.
\end{split}
\end{equation} 

Although Equation \ref{eq:e4functional} involves fully linked diagrams that are connected overall, there may be instances where the underlying diagram is internally disconnected and therefore requires cancellation. The Supplementary Material discusses how to resolve such examples, and also provides the complete derivations for UCC(2), UCC(3), and UCC(4) in the case of non-canonical HF orbitals.

With this in mind, several internal cancellations are found in Equation \ref{eq:e4functional}, leading to a simplified expression for the UCC(4) functional. From there, the residual equations can be formulated:

\begin{subequations}\label{eq:UCC4residEqns}
        \begin{align}
     \frac{ \partial\Delta E(4)}{\partial T_1^\dagger} = 0 \Rightarrow &D_1T_1=W_N + W_NT_2 + W_NT_1 + \frac{1}{2}\big(\frac{1}{2}W_NT_1^2+T_1^{\dagger}W_NT_1\big)  +T_2^{\dagger}W_NT_2,\\
     \frac{ \partial\Delta E(4)}{\partial T_2^\dagger} = 0 \Rightarrow &D_2T_2=W_N + W_NT_2 + W_NT_1 + \frac{1}{2}\big(\frac{1}{2}W_NT_2^2+T_2^{\dagger}W_NT_2\big) +W_NT_3+T_1^{\dagger}W_NT_2+W_NT_1T_2, \\
     \frac{ \partial\Delta E(4)}{\partial T_3^\dagger} = 0 \Rightarrow &D_3T_3=W_NT_2.
    \end{align}
\end{subequations}
After inserting these stationary conditions into the simplified form of Equation \ref{eq:e4functional}, the final, reduced UCC(4) energy is shown to be
\begin{equation}\label{eq:finalUCC4eqn}
   \Delta E(4) = \braket{0|W_NT_2|0} +\braket{0|W_NT_1|0} - \frac{1}{4}\bigg(\braket{0|(T_1^{\dagger})^2W_NT_1|0} + \braket{0|(T_2^{\dagger})^2W_NT_2|0}\bigg).
\end{equation}

 The following developments focus on Equations \ref{eq:UCC4residEqns}  and \ref{eq:finalUCC4eqn} where we are only interested in fully iterating the singles/doubles equations. In order to derive perturbative corrections designed to account for missing $T_3$-like excitations in UCCSD-like methods, we ``trace" the residual equations - starting with  the $T_3$ equation of Equation \ref{eq:UCC4residEqns} - to determine this operator's role in the UCC(4) energy. As the derived UCC(4) equations are subsumed within those of UCC($n\rightarrow \infty$), any set of $T_3$ corrections designed for UCCSD(4) are equally viable for infinite-order UCCSD.

\begin{figure}[ht!]
\centering
\includegraphics[width=\columnwidth]{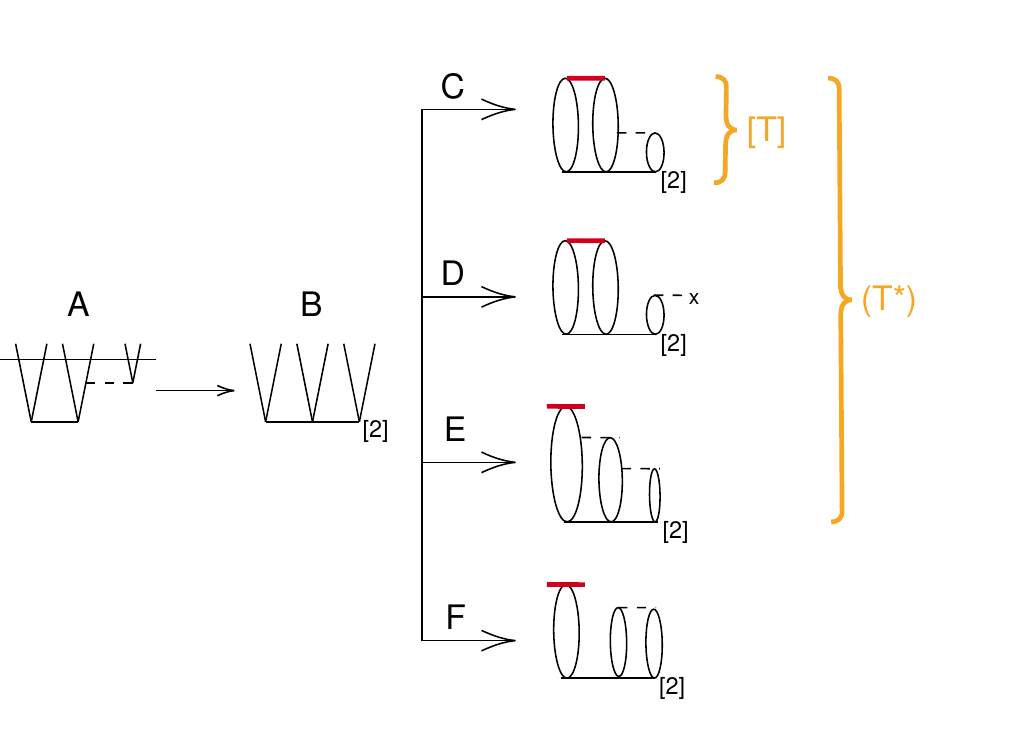}    

\caption{Outline of the procedure to extract triples' energy corrections by A) approximating $T_3$ with $(W_NT_2)_C$, B) we now have a $T_3$ correct through second-order in MBPT. If the UCC(4) equations are traced, diagram C) - which completely defines the [T] correction - and D) are rigorously shown to originate at fourth-order in MBPT. These diagrams, along with diagram E) - which represents a fifth-order contribution in MBPT - define the (T$^*$) method. The standard (T) formulation, which is composed of diagrams C), D) and F), is also studied in this work, despite the fact that diagram F) does not exist in UCC(4).}
\label{fig:squarebrackT}
\end{figure}

This procedure starts by circumnavigating the explicit solution for the $T_3$ equations by adopting the approximation
\begin{equation}\label{eq:t3info}
\begin{split}
    T_3^{[2]}&=\frac{1}{D_3}(W_NT_2)_C \\
    \end{split}
\end{equation}
using $T_2$ amplitudes from any converged, UCCSD-like calculation. The superscript denotes the order in MBPT through this contribution is correct, meaning that in this case, $T_3^{[2]}$ is correct through second-order in MBPT. Although $T_3$ is not directly specified in Equation \ref{eq:finalUCC4eqn}, it does couple to the $T_2$ equation. Inserting Equation \ref{eq:t3info} into \ref{eq:UCC4residEqns}b, we find that
\begin{equation}\label{eq:approxT2solns}
\begin{split}
    T_2^{[3]}&\equiv \frac{1}{D_2}(W_NT_3^{[2]} + f_{ia}T_3^{[2]})_C, \\
\end{split}
\end{equation}
 Note that these contributions are ``new" in the sense that they originate from an approximate solution to $T_3$. By inserting this ``new" $T_2^{[3]}$ into the first term of Equation \ref{eq:finalUCC4eqn}, we can recover a correction to either UCCSD(4) and/or the infinite-order UCCSD energy which solely originates from $T_3^{[2]}$
\begin{equation}\label{eq:sbt}
\begin{split}
    \Delta E(T_3^{[2]}) &= \braket{0|W_N T_2^{[3]}|0} \\
                         &\approx \braket{0|T_2^{\dagger}\bigg(W_NT_3^{[2]} + f_{ia}T_3^{[2]}\bigg)_C|0}.
\end{split}
\end{equation}
Note that the standard convention is to cap with an infinite-order $T_2^{\dagger}$, shown in the final line of Equation \ref{eq:sbt}. These two terms ultimately lead to diagrams C and D in Figure \ref{fig:squarebrackT}, where diagram C in particular defines the [T] correction\cite{urban1985towards} and the combination of diagrams C, D, and F account for the standard (T) correction for non-canonical orbitals. 

The remaining diagram in (T) takes the form $\braket{0|T_1^{\dagger}W_NT_3^{[2]}|0}$ and appears as F in Figure \ref{fig:squarebrackT}. As shown in the Supplementary Material, this term is completely canceled in the UCC(4) functional. Consequently, this diagram is not rigorously achievable by beginning with $T_3$ and tracing the residual equations.

However, a similar term of the form $Q_1\big(T_2^{\dagger}(W_NT_2)_C\big)$ does exist in the UCC(4) $T_1$ equation, where the subtle relationship with the $T_3$ equations is noted and $Q_1$ is the projector onto single excitations.  Inserting this term directly into the UCC(4) energy expression gives rise to a diagram that is strikingly similar to diagram F, but which is independent of the set of triples excitations that are directly tied to the $T_3$ operator; this can be seen by the absence of any factors of $\frac{1}{D_3}$. Despite yielding a net excitation effect that \textit{appears} as triples, the overall diagram is managed solely by products of  $T_1$, $T_2$, and $W_N$ making its contribution redundant  for our purposes. 

We can nevertheless search for a similar diagram that caps with a $T_1^{\dagger}$, by tracing the logic followed by the [T] developments. Assuming we are interested in only low-order contributions, we can start by solving the $T_3$ residual equations of Equation \ref{eq:UCC4residEqns} for second-order $T_3$:
    
\begin{equation}
\label{eq:t32}
  T_3^{[2]}=\frac{1}{D_3}(W_NT_2)_C,
\end{equation}
then plug Equation \ref{eq:t32} to determine a contribution to Equation \ref{eq:UCC4residEqns}b:
\begin{equation}
    T_2^{[3]}=\frac{1}{D_2}\bigg(W_N\big(\frac{1}{D_3}(W_NT_2)_C\big)\bigg)_C,
\end{equation}
which in turn couples to \ref{eq:UCC4residEqns}a:
    \begin{equation}
    \label{eq:t14}
        T_1^{[4]}=\frac{1}{D_1}\Bigg(W_N \frac{1}{D_2}\bigg(W_N\big(\frac{1}{D_3}(W_NT_2)_C\big)\bigg)_C \Bigg)_C,
    \end{equation}
and finally, we insert Equation \ref{eq:t14} into Equation \ref{eq:finalUCC4eqn} to determine what amounts to a fifth-order correction to the energy that also originates from $T_3$
    \begin{equation}\label{eq:PTdiagram}
    \begin{split}
        \Delta E(T_1^{[4]})&=\braket{0|W_NT_1^{[4]}|0}\\
        &\approx \braket{0|T_1^{\dagger}\Bigg(W_N \frac{1}{D_2}\bigg(W_N\big(\frac{1}{D_3}(W_NT_2)_C\big)\bigg)_C \Bigg)_C|0},
        \end{split}
    \end{equation}

The combination of the last line of Equation \ref{eq:PTdiagram} - depicted in  Figure \ref{fig:squarebrackT} E - and Equation \ref{eq:sbt} leads to a new method referred to as (T$^*$).

As before, we choose to cap with an infinite-order $T_1^{\dagger}$ in Equation \ref{eq:PTdiagram} instead of $T_1^{[1]\dagger}$ to define the correction.

The following UCC results are obtained using the XACC quantum computing framework,\cite{mccaskey2020xacc} and PySCF\cite{sun2018pyscf} to generate the Hamiltonians, to calculate FCI energies, and to select important $\tau$'s suggested by CCSD amplitudes. Converged UCC amplitudes are then manipulated by the UT2 python module to extract the triples corrections. Standard CC calculations were performed using CFOUR\cite{matthews2020coupled} and ACESII\cite{perera2020advanced}. The STO-6G\cite{Hehre1969} basis set was used throughout this work, and core orbitals were dropped, with details regarding the electronic state and geometries being provided in Table S1 of the Supplementary Material. 

All FCI, UCCSD, UCCSD[T], and UCCSD(T$^*$) energies are provided in Table S2 of the Supplementary Material. Despite the lack of a rigorous basis, we also provide results using the (T) correction - denoted as UCCSD(T).

With quantum computing in the background, two forms of UCC ansatz are adopted here. The first follows closely the standard CC formalism where we simply replace the routine $T$ cluster operator by its anti-Hermitian analog $\tau$ and construct the exponential wave operator:

\begin{equation}
    %e^{\tau_1 + \tau_2}
    \label{eq:uccsd}
    |\Psi_\text{UCCSD}\rangle = e^{\sum_{ia}\theta_i^a (a^\dagger i - \text{h.c.}) + \sum_{ijab}\theta_{ij}^{ab}(a^\dagger b^\dagger ij- \text{h.c.})} |0\rangle.
\end{equation}

As previously noted, this ansatz does not naturally truncate the underlying Schr\"{o}dinger equation as does the standard $T$. A form of this operator more suitable for implementation on a quantum computer is the Trotterized or disentangled form of the ansatz, which we refer to as tUCCSD hereafter

\iffalse
\begin{equation}
    %\prod_{k}e^{\theta_k \big(T_{1k}-T_{1k}^{\dagger}\big)}\prod_{k}e^{\theta_k \big(T_{2k}-T_{2k}^{\dagger}\big)}
     |\Psi_\text{tUCCSD}\rangle = \prod_{ia}e^{\theta_i^a(a^\dagger i- \text{h.c.})}\prod_{ijab}e^{\theta_{ij}^{ab}(a^\dagger b^\dagger ij- \text{h.c.}) |0\rangle},
\end{equation}
\fi

\begin{equation}
    %\prod_{k}e^{\theta_k \big(T_{1k}-T_{1k}^{\dagger}\big)}\prod_{k}e^{\theta_k \big(T_{2k}-T_{2k}^{\dagger}\big)}
     |\Psi_\text{tUCCSD}\rangle = \prod_{IA}e^{\theta_I^A(A^\dagger I- \text{h.c.})}e^{\theta_{\bar{I}}^{\bar{A}}(\bar{A}^\dagger \bar{I}- \text{h.c.})} \prod_{\substack{I<J \\ A<B}}e^{\theta_{IJ}^{AB}(A^\dagger B^\dagger IJ- \text{h.c.})}e^{\theta_{\bar{I}\bar{J}}^{\bar{A}\bar{B}}(\bar{A}^\dagger \bar{B}^\dagger \bar{I}\bar{J}- \text{h.c.})} \prod_{IJAB}e^{\theta_{I\bar{J}}^{A\bar{B}}(A^\dagger \bar{B}^\dagger I\bar{J} - \text{h.c.})} |0\rangle,
    \label{eq:ansatz}
\end{equation}
with $I, J, A, B$ indexing $\alpha$ orbitals and $\bar{I}, \bar{J}, \bar{A}, \bar{B}$ indexing the corresponding $\beta$ orbitals. 

While the UCCSD ansatz in Equation \ref{eq:uccsd} is unique, the composition of the excitation operators alone will not uniquely define its Trotterized analog, since such operators do not commute in general. The potential ambiguity is removed by fully specifying the indices over the products in Equation \ref{eq:ansatz}. Such ordering is in line with previous results. However, it is not adequate for geometries away from equilibrium, as will be discussed later.\cite{grimsley2019trotterized} As shown later on, the energy difference between UCCSD and tUCCSD - as well as the corresponding triples corrections  - is nominal.

In this work we analyze the performance of three classes of CC/UCC methods:

\begin{enumerate}
    \item ``Standard" CC methods that incorporate triples corrections into CCSD, namely CCSD[T], CCSD(T), CCSDT-1, and $\Lambda$CCSD(T);
    \item Truncated UCC methods through fourth-order;
    \item Infinite-order UCCSD and tUCCSD - and the corresponding [T], (T$^*$), and (T) corrections.
\end{enumerate}

The main goal of this paper is to evaluate the performance of methods under class 3, that is, UCCSD[T]/(T)/(T$^*$) and tUCCSD[T]/(T)/(T$^*$). Those are built with the set of optimal $\tau_1$ and $\tau_2$ learned by subjecting the two choices of ans\"{a}tze to the Variational Quantum Eigensolver algorithm (VQE)\cite{Peruzzo2014} 

\begin{equation}
    \tau_1^*, \tau_2^* =  \argminA_{\tau_1, \tau_2}\braket{\Psi(\tau_1, \tau_2)| H |  \Psi(\tau_1, \tau_2)},
\end{equation}
with $\tau_1^*$ and $\tau_2^*$ being employed in the diagrams in Figure \ref{fig:squarebrackT} and Equations \ref{eq:sbt}-\ref{eq:PTdiagram}.

Table \ref{tab:errors} records the error of each method against FCI, which is found in the range of 50-271 mH, and the corresponding percentage of total correlation energy. ``Standard" CC methods largely align with FCI, with CCSD missing up to 16 mH of correlation energy in the worst case. The inclusion of (T) represents a significant improvement, where we close the diagram of Figure \ref{fig:squarebrackT} with a converged $T_2^{\dagger}$ amplitude. We note that $T_2^{\dagger}$ is the lowest-order approximation to the complete solution of the left-hand eigenvalue problem defining $\Lambda_2$.\cite{taube2009rethinking} If we, instead, cap the diagram of Figure \ref{fig:squarebrackT}  using $\Lambda_2$ - resulting in $\Lambda$CCSD(T) -  the results are slightly worse than CCSD(T), except for CO. This is somewhat counter-intuitive, but is a trend that has already been previously observed.\cite{taube2008improving} Overall, the infinite-order CCSDT-1 shows the best performance amongst the methods we consider. It takes the $T_2$ portion of the diagram in Figure \ref{fig:squarebrackT} and adds it to the CCSD $T_2$ residual equations, accounting for some coupling between $T_2$ and a diagram that originates from $T_3$ at lowest order. Thus, the energy ``feels" effects from this diagram that originate from $T_3^{[2]}$, and is responsible for this improvement. 

\begin{table}[ht]
    \centering
    \caption{Error with respect to FCI, in mH, and the corresponding percentage of the total correlation energy (in parentheses). $\ddagger$ denotes an all electron calculation.}
    \label{tab:errors}
\begin{tabular}{|c| |c |c |c |c |c|} 
 \hline
Method & H$_2$O & CO & C$_2$ & O$_2$ & N$_2$ \\ \hline
UCCSD & -0.100 (99.79) & -8.183 (94.12) & -11.90 (95.61) & -9.132 (94.15) & -2.176 (98.62) \\
UCCSD[T] & -0.023 (99.95) & -6.105 (95.61) & 4.841 (101.78) & -6.018 (96.14) & -0.383 (99.75) \\
UCCSD(T) & -0.015	(99.97)&-5.973	(95.71)&8.247	(103.04)&-5.979	(96.17)&	-0.356 (99.78)\\
UCCSD(T*)&-0.023 (99.95) &	-6.071 (95.64) &	7.004 (102.58)&	-6.002 (96.16)& 	-0.371 (99.77) \\			
tUCCSD & -0.098 (99.80) & -7.888 (94.33) & -11.08 (95.91) & -9.122 (94.16) & -2.172 (98.62) \\
tUCCSD[T] & -0.020 (99.95) & 1.776 (101.27) & 6.109 (102.25) & -5.842 (96.25) & -0.623 (99.60) \\ 
tUCCSD(T) &-0.012	(99.98)&3.021(102.17) &	9.967 (103.67)&	-5.801 (96.29)&	-0.597 (99.62) \\
tUCCSD(T*)& -0.021 (99.96) & 2.147 (101.54) &	8.565 (103.16) &-5.826 (96.27) &	-0.612 (99.61)\\
\hline
CCSD & -0.118 (99.76) & -8.157 (94.14) & -16.33 (93.98) & -10.47 (93.29) & -3.983 (97.48) \\
CCSD(T) & -0.050 (99.89) & -0.865 (99.37) & -2.817 (98.96) & -7.411 (95.25) & -2.231 (98.59) \\
CCSDT-1 & -0.048 (99.9) & 1.163 (100.83) & -2.774 (98.97) & -6.201 (96.02) & -2.164 (98.63) \\
$\Lambda$CCSD(T) & -0.085 (99.82) & 0.574 (100.41) & -4.375 (98.38) & -7.819 (94.99) & -2.294 (98.55) \\ \hline
UCC(2) & -14.22 (71.56) & -10.74 (92.28) & -25.29 (90.68) & -29.76 (80.94) & -2.921 (98.15) \\
UCC(3) & 0.338 (100.67) & -9.795 (92.96) & -14.97 (94.48) & -2.312 (98.51) & -1.036 (99.34) \\
UCC(4)$^\ddagger$ & 0.942 (101.88) & 9.555 (106.84) & 31.46 (111.56) & -21.77 (88.81) & 1.707 (101.07) \\
UCCSD(4)$^\ddagger$ & 0.858 (101.71) & 2.326 (98.33) & 6.288 (102.31) & -31.61 (83.75) & -0.346 (99.78) \\
UCCSD(4)[T]$^\ddagger$ & 0.936 (101.87) & 7.697 (105.51) & 19.81 (107.28) & -29.77 (84.7) & 1.396 (100.87) \\ \hline
%HF & -50.013 & -139.287 & -271.418 & -156.205 & -158.470 \\ \hline
\end{tabular}

\end{table}

% Start with truncated UCC trends
Moving to the truncated UCC methods, we see that UCC(2) - or equivalently MBPT(2) - captures 71-98\% of the correlation energy across the set of molecules tested, and is consistently above FCI by 3-30 mH. UCC(3) - or equivalently LCCD - is a marked improvement, reporting errors no larger than 15 mH. The results from UCC(4) are less straightforward; except in the case of CO, UCC(3) results are in overall better agreement with FCI. We also note that - except in the case of the $^1\Delta$ state of O$_2$ - UCC(4) consistently underestimates FCI. A byproduct of prematurely terminating the commutator expansion is that we forgo any guarantee of achieving an upper bound to FCI, as might otherwise be realized by a fully variational method. Noting the exception of O$_2$ again, we find that omitting the triples portion of UCC(4) - which defines the UCCSD(4) method - has been shown to lead to improvements,\cite{watts1989unitary} and is of comparable quality to UCC(3). By adding the [T] correction for the missed triple excitations on top of UCCSD(4), the resulting UCCSD(4)[T] method - with the exception being O$_2$ - seems to represent a middle ground between UCCSD(4) and UCC(4). This indicates that adding perturbative triples corrections is a step toward the complete UCC(4) method which iterates the triples residual equations. 

% Discuss the infinite-order UCC methods and corrections
Turning to the infinite-order UCC methods, both the UCCSD and tUCCSD methods are of comparable or superior quality to the standard CCSD method, and in either representation of UCCSD adding [T] yields a clear and consistent improvement over the baseline ans\"{a}tze. In fact, for H$_2$O, C$_2$, and N$_2$ the [T] variants of UCCSD are within 1\% of the FCI. For O$_2$, adding [T] nevertheless improves upon the t/UCCSD energy by 2\%. This is also true for CO when using UCCSD[T], but for this case the tUCCSD[T] result recovers 6\% to the tUCCSD correlation energy. By adding the (T) and (T$^*$) diagrams, we do not find appreciable performance gains over baseline [T]. In fact, for C$_2$ both t/UCCSD[T] models are noticeably better than their alternative triples counterparts. 

\begin{figure*}[ht]
 \centering
 \includegraphics[width=\columnwidth]{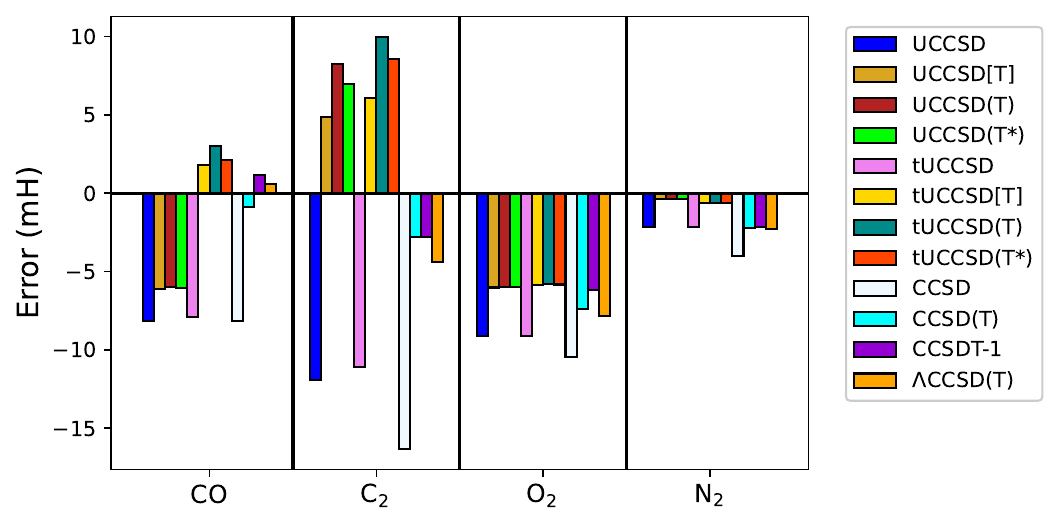}
 \caption{Bar plot of the errors reported in Table \ref{tab:errors}}
 \label{fig:fullBarplot}
\end{figure*}

% comparison between UCC CC and UCC(n)
A visual comparison of these infinite-order UCC methods and their ``standard" CC counterparts is shown in Figure \ref{fig:fullBarplot}. Here, we see that the UCC-based [T] offers the best performance for both N$_2$ and O$_2$, whereas the standard CC (T) correction yields better results for CO and C$_2$. 

\begin{figure*}[ht]
 \centering
 \includegraphics[width=\columnwidth]{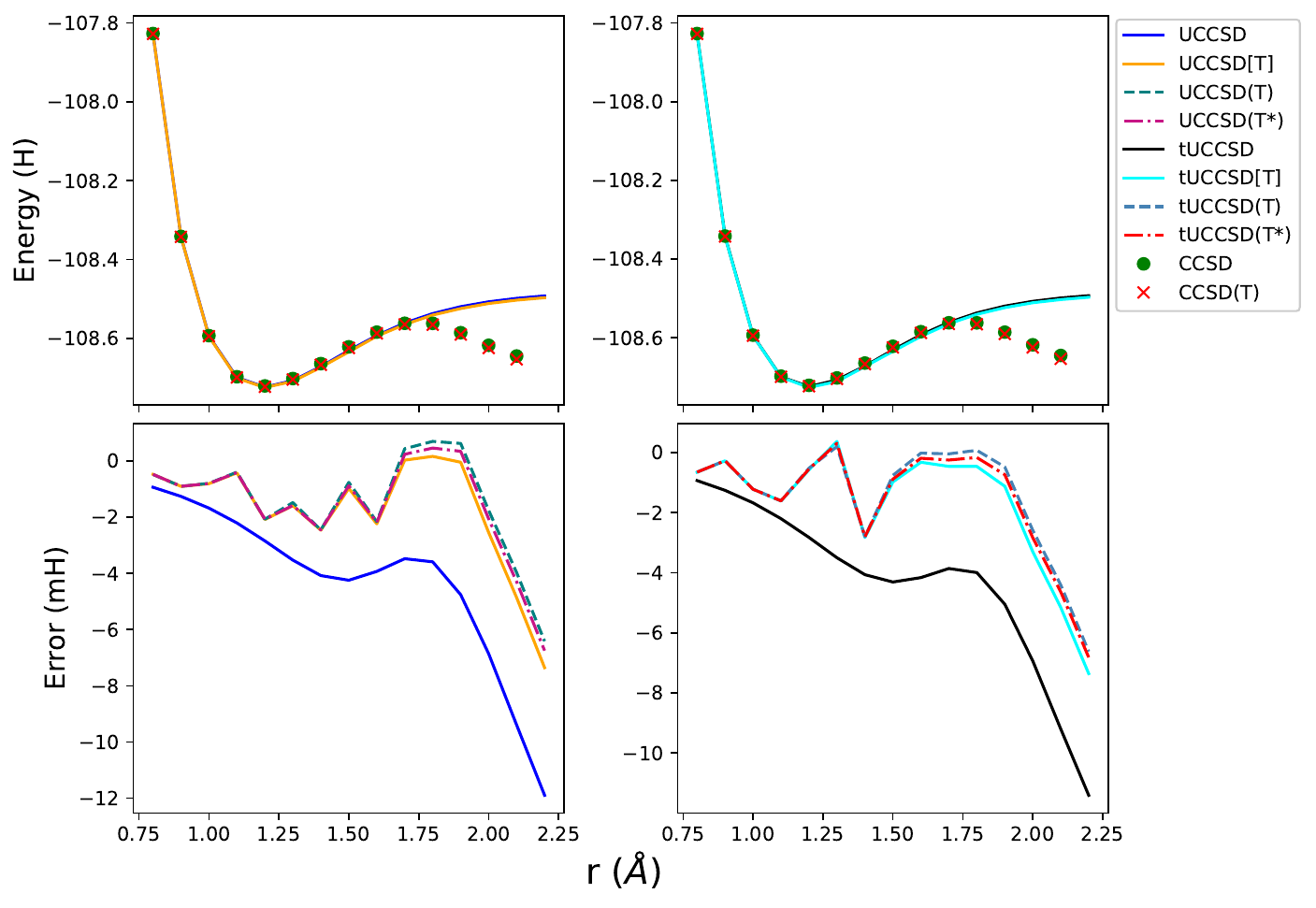}
 \caption{Comparison of the UCCSD/[T]/(T)/(T$^*$) and tUCCSD/[T]/(T)/(T$^*$) results for the dissociation of N$_2$. }
 \label{fig:UCCSDN2}
\end{figure*}

Outside the equilibrium region, the [T] correction to UCCSD yields more dramatic improvements as compared to the standard CCSD and CCSD(T) methods. This is illustrated in Figure \ref{fig:UCCSDN2} which show the potential energy surface (PES) of N$_2$. Around 1.75 \AA, standard CCSD/CCSD(T) methods begin to display the typical drastic divergence from FCI. However, both variants of UCCSD[T] results are better-behaved in this regime. 

Beyond 2.2 \AA, tUCCSD[T] exhibits perplexing behavior that is not seen in UCCSD[T], as shown in Figure S1 of the Supplementary Material. It is known that one byproduct of the tUCCSD ansatz is the inherent sensitivity to operator ordering.\cite{grimsley2019trotterized} While both UCCSD and tUCCSD ans\"{a}tze remain variational upper bounds to FCI - as expected - it is intriguing that only the [T] corrections built upon tUCCSD behave poorly in this region. Further analysis of the role of the operator ordering is deferred to the Supplementary Material.

%\clearpage

%\section{Conclusion}

In summary, we study finite-orders of UCC theory to design a perturbative treatment for triple excitations in infinite-order UCCSD, in line with what gave rise to the so-called ``gold standard'' of quantum chemistry. We show that such an approach reliably improves the energy of several small molecules at - or near - equilibrium geometries as compared to baseline UCCSD. This result is independent of whether the full or Trotterized UCCSD operator is used. We find that the differences between the (T) and (T$^*$) approaches are largely indistinguishable, and do not represent significant improvements over the [T] results. An important finding is that the corrections presented here are much more resilient to the typical breakdown characteristic of methods based on perturbation theory when we go beyond the Coulson-Fischer point, in this case for N$_2$, despite the tUCCSD-based corrections eventually displaying erroneous behavior which can be attributed to the operator ordering in the tUCCSD ans\"{a}tze.

The current work opens the door to several topics worth exploring. The most immediate is that it naturally lends itself to the development of additional perturbative corrections. For example, it is conceivable that energy corrections originating from $T_1$ could be constructed to account for missing single excitation effects in infinite-order t/UCCD by studying the UCC(4) equations for $T_1$ and tracing the logic followed in the development of UCCSD[T]. In addition to the inherit formal contribution of MBPT in conjunction with the UCC ansatz, the ideas presented here have potential applications in quantum computing. More specifically, energy corrections become accessible from hybrid algorithms -- e.g. VQE --  with a classical post-processing step free from requiring extra quantum resources. This is embodied by the evaluation of the diagrams in Figure \ref{fig:squarebrackT} constructed using $\tau_1$ and $\tau_2$, which can be obtained on a quantum computer. Continued efforts focusing on the symbiotic relationship between MBPT and UCC that exploit hybrid classical/quantum computing algorithm paradigms will be the target of forthcoming work.

\section*{Acknowledgements}
D.C. and Z.W.W. thank Karol Kowalski for various fruitful discussions. Z.W.W also thanks Dr. Taylor Barnes of MolSSI for his guidance in developing the UT2 software. This work was supported by the Air Force Office of Scientific Research under AFOSR Award No. FA9550-23-1-0118. Z.W.W. thanks the National Science Foundation and the Molecular Sciences Software Institute for financial support under Grant No. CHE-2136142. Z.W.W. also acknowledges support from the U.S. Department of Energy, Office of Science, Office of Workforce Development for Teachers and Scientists, Office of Science Graduate Student Research (SCGSR) program. The SCGSR program is administered by the Oak Ridge Institute for Science and Education (ORISE) for the DOE. ORISE is managed by ORAU under contract number DE-SC0014664. D.C. acknowledges support by the “Embedding Quantum Computing into Many-body Frameworks for Strongly Correlated Molecular and Materials Systems” project, which is funded by the U.S. Department of Energy (DOE), Office of Science, Office of Basic Energy Sciences, the Division of Chemical Sciences, Geosciences, and Biosciences.

\bibliographystyle{iopart-num}
\bibliography{main.bib}

%\clearpage

%\includepdf[pages={{},-},width=\textwidth]{si.pdf}

\end{document}

\title{Supplementary Material for A new ``gold standard'': perturbative triples corrections in unitary coupled cluster theory and prospects for quantum computing}

\thanks{This manuscript has been authored by UT-Battelle, LLC, under Contract No.~DE-AC0500OR22725 with the U.S.~Department of Energy. The United States Government retains and the publisher, by accepting the article for publication, acknowledges that the United States Government retains a non-exclusive, paid-up, irrevocable, world-wide license to publish or reproduce the published form of this manuscript, or allow others to do so, for the United States Government purposes. The Department of Energy will provide public access to these results of federally sponsored research in accordance with the DOE Public Access Plan.}
%\date{}
\author{Zachary W. Windom$^{1, 2}$, Daniel Claudino$^2$\footnote{\href{mailto:claudinodc@ornl.gov}{claudinodc@ornl.gov}}, and Rodney J. Bartlett$^1$}
\affiliation{$^1$Quantum Theory Project, University of Florida, Gainesville, FL, 32611, USA \\
$^2$Computational Sciences and Engineering Division,\ Oak\ Ridge\ National\ Laboratory,\ Oak\ Ridge,\ TN,\ 37831,\ USA}

\maketitle

\section{Supporting derivations of finite-order UCC equations}
\renewcommand\theequation{S\arabic{equation}}
\renewcommand\thefigure{S\arabic{figure}}
\renewcommand\thetable{S\arabic{table}}  

In the case where an ambiguous expression could conceivably have internally disconnected (D) pieces, the overall connected (and necessarily linked) diagram is decomposed into underlying connected (C) and disconnected (D) pieces using
\begin{equation}\label{eq:decomposeDiagrams}
   \braket{0|XH_NY|0}_C = \frac{1}{2}\bigg(\braket{0|X\big(H_NY\big)_D + X\big(H_NY\big)_C + \big(XH_N\big)_DY + \big(XH_N\big)_CY|0} \bigg) 
\end{equation}for $X$ and $Y$ non-vanishing arbitrary strings of $T_n$ and $T_n^{\dagger}$ operators, and the factor of $\frac{1}{2}$ compensating for over-counting. The internally disconnected (D) pieces can be factored into a connected (C) form via
\begin{equation}\label{eq:factorTn}
    T_n^{\dagger}\bigg(H_NT_iT_j\bigg)_D =T_n^{\dagger}T_i\bigg(H_NT_j\bigg)_C + T_n^{\dagger}T_j\bigg( H_NT_i\bigg)_C
\end{equation} 

As the following derivations define the UCC energy functional with respect to various MBPT orders, we follow the convention that $\Delta E^{[i]}$ indicates a correction at a particular order, $i$, all of which can be summed through order $n$ to define the cumulative correction, $\Delta E(n)$:
\begin{equation}\label{eq:cumulativeE}
    \Delta E(n) = \sum_{i=2}^{n} \Delta E^{[i]},
\end{equation} Furthermore, we assume the non-canonical orbitals have been semi-canonicalized such that $f_{ij}=f_{ab}=0$ for $i\neq j$ and $a\neq b$.

\subsection{Background for deriving the non-canonical HF UCC equations}

\subsection{Derivation of UCC(2)}

The second order energy functional is of the form
\begin{equation}\label{eq:mbpt2origenergy}
\begin{split}
    \Delta E(2)&= \braket{0|\tau_2^{\dagger}W_N|0} + \braket{0|W_N\tau_2|0} + \braket{0|\tau_2^{\dagger}f_N\tau_2|0} + \braket{0|\tau_1^{\dagger}W_N|0} + \braket{0|W_N\tau_1|0}+\braket{0|\tau_1^{\dagger}f_N\tau_1|0}\\
    &=\braket{0|T_2^{\dagger}W_N|0}+ \braket{0|W_NT_2|0} + \braket{0|T_2^{\dagger}f_NT_2|0}+\braket{0|T_1^{\dagger}W_N|0}+\braket{0|W_NT_1|0}+\braket{0|T_1^{\dagger}f_NT_1|0}
    \end{split}
\end{equation}
Upon variation of the above with respect to $T_2$ (or, equivalently, $T_2^{\dagger}$) we see that 
\begin{equation}\label{eq:mbpt2eqns}
\begin{split}
    \frac{\partial \Delta E(2)}{\partial \big(t_{ij}^{ab}\big)^{\dagger}} &= \braket{\Phi_{ij}^{ab}|\bigg( W_N + f_NT_2\bigg)|0}\\
& \implies D_2T_2= W_N,    
\end{split}
\end{equation} with the last line representing the MBPT(2) residual equations if using canonical HF orbitals. Variation with respect to $T_1^{\dagger}$ leads to 
\begin{equation}\label{eq:mppt2t1resids}
\begin{split}
    \frac{\partial \Delta E(2)}{\partial \big(t_{i}^{a}\big)^{\dagger}} &=\braket{\Phi_i^a|\bigg(f_{ai}+f_NT_1\bigg)|0} \\
    & \implies D_1T_1=f_{ai}
\end{split}
\end{equation}
By inserting the residual condition found in Equations \ref{eq:mbpt2eqns} and \ref{eq:mppt2t1resids} into Equation \ref{eq:mbpt2origenergy} and further invoking the resolution of the identity - $1 = \ket{0}\bra{0} + \ket{\Phi_{ij}^{ab}}\bra{\Phi_{ij}^{ab}}$ - the final energy expression becomes the familiar one from MBPT(2) for general, non-canonical HF references
\begin{equation}
    \Delta E(2)=\braket{0|W_NT_2|0} + \braket{0|f_{ia}T_1|0}
\end{equation}

\subsection{Derivation of UCC(3)}
At third-order, we are interested in the functional 
\begin{equation}\label{eq:UCC3origEqns}
\begin{split}
    \Delta E^{[3]}=&\bigg(-\frac{1}{2}\braket{0|T_2^{\dagger}T_1f_NT_1 + \text{h.c.}|0} +\frac{1}{2}\braket{0|(T_1^{\dagger})^2f_NT_2+h.c|0}\bigg) \\
    &+\bigg(-\frac{1}{2}\braket{0|T_2^{\dagger}T_1W_N+\text{h.c.}|0}+\frac{1}{2}\braket{0|(T_1^{\dagger})^2W_N+\text{h.c.}|0}\bigg) \\
    &+\braket{0|(T_1^{\dagger}+T_2^{\dagger})W_N(T_1+T_2)|0}
\end{split}
\end{equation}
We will show that the first two lines in the above completely cancel, leaving the final term to contribute in the UCC(3) energy functional. To show this, we will expand those expressions and group relevant terms such that
\begin{equation}\label{eq:UCC3invoke}
\begin{split}
   & -\bigg(\braket{0|T_2^{\dagger}T_1f_NT_1|0} + \text{h.c.}\bigg) \equiv \braket{0|T_2^{\dagger}T_1W_N|0}+h.c \\
   & -\bigg(\braket{0|\big(T_1^{\dagger}\big)^2f_NT_2|0} + \text{h.c.}\bigg) \equiv \braket{0|\big(T_1^{\dagger}\big)^2W_N|0}+h.c\\
\end{split}
\end{equation} where strict equivalence is found once we arrange the l.h.s. and r.h.s. to emulate the UCC(2) residual equations. This exhausts the viable contractions in the first two terms of Equation \ref{eq:UCC3origEqns}, leaving only the last term in the energy functional. 

Thus, the cumulative functional at third order can be expressed as
\begin{equation}\label{eq:UCC3cumulativeEqns}
\begin{split}
    \Delta E(3)&=\underbrace{\braket{0|\bigg(T_1^{\dagger}+T_2^{\dagger}\bigg)W_N\bigg(T_1+T_2\bigg)|0}}_{\text{$\Delta E^{[3]}$}} \\
    &+\underbrace{\braket{0|T_2^{\dagger}W_N|0}+ \braket{0|W_NT_2|0} + \braket{0|T_2^{\dagger}f_NT_2|0}+\braket{0|T_1^{\dagger}W_N|0}+\braket{0|W_NT_1|0}+\braket{0|T_1^{\dagger}f_NT_1|0}}_{\text{$\Delta E^{[2]}$}}
\end{split}
\end{equation}

Determining the corresponding set of residual equations requires variation of the above with respect to both $T_2$/$T_2^{\dagger}$ and $T_1$/$T_1^{\dagger}$,  leading to
\begin{equation}
    \begin{split}
    \frac{\partial \Delta E(3)}{\partial \big(t_{ij}^{ab}\big)^{\dagger}} &=\braket{\Phi_{ij}^{ab}|\bigg( W_N + f_NT_2 + W_NT_2 + W_NT_1\bigg)|0}\\
    &\implies D_2T_2=W_N + W_NT_2 + W_NT_1
    \end{split}
\end{equation} which defines the $T_2$ equations and 
\begin{equation}
        \begin{split}
    \frac{\partial \Delta E(3)}{\partial \big(t_{i}^{a}\big)^{\dagger}} &=\braket{\Phi_i^a|\bigg( f_{ai}+f_NT_1+W_NT_1+W_NT_2+T_1^{\dagger}W_N\bigg)|0} \\
    & \implies D_1T_1=f_{ai}+W_NT_1+W_NT_2\\
\end{split}
\end{equation} which defines the $T_1$ equations.

Following the procedure detailed in UCC(2), we can insert the residual conditions found above into the energy expression of Equation \ref{eq:UCC3cumulativeEqns} which yields the final expression for the third-order energy
\begin{equation}
    \Delta E(3) = \braket{0|W_NT_2|0} +\braket{0|f_{ia}T_1|0}
\end{equation}The set of equations generated by this procedure are the linearized CCSD equations for non-HF examples. 

\subsection{Derivation of UCC(4)}
Turning to the fourth-order energy functional:
\begin{equation}
\begin{split}
\Delta E^{[4]}=\Delta E^{[4A]}+\Delta E^{[4B]}
%    \Delta E^{[4]} &= \underbrace{\bigg( \braket{\Phi^{(1)}|f_N|\Phi^{(3)}|0} + \text{h.c.}\bigg) + \bigg(\braket{0|W_N|\Phi^{(3)}}+ \text{h.c.}\bigg) }_{\text{\textbf{A}}} + \\
%    &+\underbrace{\braket{\Phi^{(2)}|f_N|\Phi^{(2)}}+\bigg(\braket{\Phi^{1}|W_N|\Phi^{(2)}}+h.c\bigg)}_{\textbf{B}},
\end{split}
\end{equation}
where
\begin{equation}
    \begin{split}
        \Delta E^{[4A]}&=-\braket{0|\frac{1}{3!}\bigg((T_1^{\dagger})^2T_1f_NT_1+\text{h.c.}+(T_2^{\dagger})^2T_2f_NT_2+\text{h.c.}\bigg)+\frac{1}{2}\bigg(T_3^{\dagger}T_1f_NT_2+\text{h.c.}+T_3^{\dagger}T_2f_NT_1+\text{h.c.}\bigg)|0}\\
        &-\frac{1}{3}\braket{0|T_1^{\dagger}T_2^{\dagger}T_1f_NT_2+\text{h.c.}+T_2^{\dagger}T_1^{\dagger}T_2f_NT_1+\text{h.c.}|0}\\
        &-\braket{0|\frac{1}{3!}\Big((T_1^{\dagger})^2T_1W_N+\text{h.c.}\Big) +\frac{1}{3!}\Big((T_2^{\dagger})^2T_2W_N+\text{h.c.}\Big)+\frac{1}{2}\Big(T_3^{\dagger}T_1W_N+\text{h.c.}+T_3^{\dagger}T_2W_N+\text{h.c.}\Big)|0} \\
        &-\frac{1}{3}\braket{0|T_1^{\dagger}T_2^{\dagger}T_1W_N+\text{h.c.}+T_2^{\dagger}T_1^{\dagger}T_2W_N+\text{h.c.}|0}-\frac{1}{6}\braket{0|T_2^{\dagger}T_1T_1^{\dagger}f_NT_2+\text{h.c.}+T_2^{\dagger}T_1T_1^{\dagger}W_N+\text{h.c.}|0}
    \end{split}
\end{equation}

\begin{equation}
    \begin{split}
        \Delta E^{[4B]}&=\frac{1}{4}\braket{0|(T_1^{\dagger})^2f_NT_1^2+(T_2^{\dagger})^2f_NT_2^2|0}+\braket{0|T_3^{\dagger}f_NT_3|0}+\braket{0|T_3^{\dagger}f_NT_1T_2+\text{h.c.}|0}+\braket{0|T_1^{\dagger}T_2^{\dagger}f_NT_1T_2|0}\\
        &+\frac{1}{2}\braket{0|(T_1^{\dagger})^2W_NT_1+\text{h.c.}+(T_2^{\dagger})^2W_NT_2+\text{h.c.}|0}+\braket{0|T_3^{\dagger}W_N(T_1+T_2)+\text{h.c.}|0}+\braket{0|T_2^{\dagger}T_1^{\dagger}W_N(T_1+T_2)+\text{h.c.}|0}\\
        &-\frac{1}{2}\braket{0|T_2^{\dagger}T_1W_N(T_1+T_2)+\text{h.c.}|0}+\frac{1}{2}\braket{0|(T_1^{\dagger})^2W_NT_2+\text{h.c.}|0}
    \end{split}
\end{equation}
we assign the functional with nomenclature \textbf{A} and \textbf{B} to denote relevant sections in which there is internal cancellation of disconnected diagrams. We note that the majority of elements that involve $f_N$ and do not linearly coupled with the $\tau_n^{\dagger}$/$\tau_n$ are in need of being canceled. 

Starting with \textbf{A}, there are seven internal cancellations that are facilitated by invoking the UCC(2) residual equations, causing all terms to cancel
\begin{equation}
    \begin{split}
        &-\bigg( T_3^{\dagger}T_1f_NT_2+\text{h.c.}\bigg)=T_3^{\dagger}T_1W_N+\text{h.c.} \\
        &-\bigg(T_3^{\dagger}T_2f_NT_1+\text{h.c.}\bigg)=T_3^{\dagger}T_2W_N+h.c \\
        & -\bigg(\big(T_1^{\dagger}\big)^2T_1f_NT_1+h.c \bigg)= \big(T_1^{\dagger}\big)^2T_1W_N+\text{h.c.} \\
        &-\bigg(\big(T_2^{\dagger}\big)^2T_2f_NT_2 +\text{h.c.}\bigg)= \big(T_2^{\dagger}\big)^2T_2W_N+\text{h.c.}\\
        &-\bigg(T_1^{\dagger}T_2^{\dagger}T_2f_NT_1+\text{h.c.}\bigg)=T_1^{\dagger}T_2^{\dagger}T_2W_N + \text{h.c.} \\
        &-\bigg(T_2^{\dagger}T_1^{\dagger}T_1f_NT_2+\text{h.c.} \bigg)= T_2^{\dagger}T_1^{\dagger}T_1W_N + \text{h.c.} \\
        &-\bigg(T_2^{\dagger}T_1T_1^{\dagger}f_NT_2+\text{h.c.}\bigg)=T_2^{\dagger}T_1T_1^{\dagger}W_N+\text{h.c.}\\
    \end{split}
\end{equation} such that $\Delta E^{[4A]}=0$.
Moving to \textbf{B} in $\Delta E^{[4B]}$, we invoke Equations \ref{eq:decomposeDiagrams} and \ref{eq:factorTn} to separate out the internally connected and disconnected pieces of $\Delta E^{[4B]}$:
\begin{equation}\label{eq:identity}
\begin{split}
    T_3^{\dagger}(f_NT_1T_2)_D =& \bigg( T_3^{\dagger}T_1\big(f_NT_2\big)_C + T_3^{\dagger}T_2\big(f_NT_1\big)_C \bigg)\\
    T_1^{\dagger}T_2^{\dagger}(f_NT_1T_2)_D=&T_1^{\dagger}T_2^{\dagger}T_2(f_NT_1)_C+T_1^{\dagger}T_2^{\dagger}T_1(f_NT_2)_C\\
    (T_1^{\dagger}T_2^{\dagger}W_NT_2)_C=&\frac{1}{2}\bigg(T_1^{\dagger}T_2^{\dagger}(W_NT_2)_D+T_1^{\dagger}T_2^{\dagger}(W_NT_2)_C\bigg)\\
    =&\frac{1}{2}\bigg(T_1^{\dagger}T_2^{\dagger}T_2(W_N)_C+T_1^{\dagger}T_2^{\dagger}(W_NT_2)_C\bigg)\\
    (T_3^{\dagger}W_NT_2)_C=&\frac{1}{2}\bigg(T_3^{\dagger}(W_NT_2)_D+T_3^{\dagger}(W_NT_2)_C \bigg)\\
    =&\frac{1}{2}\bigg(T_3^{\dagger}T_2(W_N)_C+T_3^{\dagger}(W_NT_2)_C\bigg)
\end{split}
\end{equation} Using these identities, the following internal cancellations arise in $\Delta E^{[4B]}$:
\begin{subequations}
\begin{align}
        &-\frac{1}{4}\big(T_1^{\dagger}\big)^2f_NT_1^{2} \equiv \frac{1}{4}\bigg( T_1^{\dagger}W_NT_1^2 + \text{h.c.} \bigg) \label{eqn:line-1}\\
        &-\frac{1}{4}\big(T_2^{\dagger}\big)^2f_NT_2^{2} \equiv \frac{1}{4}\bigg( T_2^{\dagger}W_NT_2^2 + \text{h.c.} \bigg) \label{eqn:line-2}\\
        &-T_1^{\dagger}T_2^{\dagger}T_2f_NT_1=T_1^{\dagger}T_2^{\dagger}W_NT_2+\text{h.c.}\label{eqn:line-3}\\
        &-T_1^{\dagger}T_2^{\dagger}T_1f_NT_2=T_1^{\dagger}T_2^{\dagger}W_NT_1+\text{h.c.}\label{eqn:line-4}\\
        &-T_3^{\dagger}T_2f_NT_1+\text{h.c.}=T_3^{\dagger}T_2W_N+\text{h.c.}\label{eqn:line-5}\\
        &-T_3^{\dagger}T_1f_NT_2+\text{h.c.}=T_3^{\dagger}T_1W_N+\text{h.c.}\label{eqn:line-6}\\
        &-\bigg(T_2^{\dagger}T_1f_NT_1+h.c\bigg)=\bigg(T_2^{\dagger}T_1W_N+\text{h.c.}\bigg)+\bigg(T_2^{\dagger}T_1W_NT_1+\text{h.c.}\bigg)+\bigg(T_2^{\dagger}T_1W_NT_2+\text{h.c.}\bigg)\label{eqn:line-7}\\
        &-\bigg((T_1^{\dagger})^2f_NT_2+h.c\bigg)=\bigg((T_1^{\dagger})^2W_N+\text{h.c.}\bigg)+\bigg((T_1^{\dagger})^2W_NT_1+\text{h.c.}\bigg)+\bigg((T_1^{\dagger})^2W_NT_2+\text{h.c.}\bigg)\label{eqn:line-8}
\end{align}
\label{eqn:all-lines}
\end{subequations}

Equations \ref{eqn:line-1} and \ref{eqn:line-2} cancel incompletely, leaving $\frac{1}{4}\bigg(\big(T_1^{\dagger}W_NT_1^{2} +\text{h.c.}\big) +T_2^{\dagger}W_NT_2^2+h.c\bigg)$ in the UCC(4) functional. The identities of Equation \ref{eq:identity} are used to invoke the cancellations found in Equations \ref{eqn:line-3}-\ref{eqn:line-6}, noting that the connected portions of $\bigg(T_1^{\dagger}T_2^{\dagger}(W_NT_2)_C+\text{h.c.}\bigg)$ and $\bigg( T_3^{\dagger}W_NT_2+\text{h.c.}\bigg)$ survive in the UCC(4) functional. Finally, note that Equations \ref{eqn:line-7} and \ref{eqn:line-8} invoke the UCC(3) $T_1$ and $T_2$ residual equations, respectively, to instigate cancellation in the UCC(4) functional by taking advantage of Equation \ref{eq:UCC3invoke}.

These simplifications lead to a signficantly reduced expression for the UCC(4) energy functional:
\begin{equation}\label{eq:completeUCC4}
\begin{split}
    \Delta E^{[4]} &= \braket{0|T_3^{\dagger}f_NT_3|0} +\frac{1}{4}\bigg(\braket{0|T_1^{\dagger}W_NT_1^2|0} +\text{h.c.} \bigg) + \frac{1}{4}\bigg(\braket{0|T_2^{\dagger}W_NT_2^2|0}+\text{h.c.}\bigg) \\
    & + \braket{0|T_3^{\dagger}W_NT_2|0} + \braket{0|T_2^{\dagger}W_NT_3|0}  + \braket{0|T_2^{\dagger}T_1^{\dagger}W_NT_2+T_2^{\dagger}W_NT_1T_2|0}
\end{split}
\end{equation}

Taking the partial derivative of the full $\Delta E(4)\equiv \Delta E^{[2]}+E^{[3]}+E^{[4]}$ functional yields the corresponding set of residual equations for UCC(4)
\begin{equation}\label{eq:UCC4residEqns}
        \begin{split}
     &D_1T_1=W_N + W_NT_2 + W_NT_1 + \frac{1}{2}\big(\frac{1}{2}W_NT_1^2+T_1^{\dagger}W_NT_1\big)  +T_2^{\dagger}W_NT_2\\
     &D_2T_2=W_N + W_NT_2 + W_NT_1 + \frac{1}{2}\big(\frac{1}{2}W_NT_2^2+T_2^{\dagger}W_NT_2\big) +W_NT_3+T_1^{\dagger}W_NT_2+W_NT_1T_2 \\
     &D_3T_3=W_NT_2
    \end{split}
\end{equation} 
Inserting these residual equations back into the energy functional of Equation \ref{eq:completeUCC4}, we find the final UCC(4) energy can be expressed as
\begin{equation}\label{eq:finalUCC4eqn}
    \Delta E(4) = \braket{0|W_NT_2|0} +\braket{0|W_NT_1|0} - \frac{1}{4}\bigg(\braket{0|(T_1^{\dagger})^2W_NT_1|0} + \braket{0|(T_2^{\dagger})^2W_NT_2|0}\bigg)
\end{equation}

\section{Molecular geometries}

\begin{table}[ht!]
\begin{center}
\begin{tabular}{|c| c |c|} 
 \hline
Molecule	&State&	r (\AA) \\ \hline
H$_2$O&$^1\Sigma$&0.95785\\
CO&$^1\Sigma$&1.128\\
C$_2$&$^1\Sigma$&1.243\\
O$_2$&$^1\Delta_g$&1.2075\\
N$_2$& $^1\Sigma$ &1.098\\
 \hline
\end{tabular}
\caption{Geometry of molecules studied in this work. H$_2$O has internal angle of 104.5$^{\circ}$. Information taken from Ref. \citenum{cccdbd}}
\label{table:GeoInfo}
\end{center}
\end{table}

\section{Molecular energies}

\begin{table}[ht!]
\begin{center}
\begin{tabular}{|c| c c c c c|} 
 \hline
Method	&H$_2$O&	CO&	C$_2$&	O$_2$&		N$_2$\\ \hline
FCI	&-75.7287768	&-112.4426091	&-75.4339744	&-149.1251956	&-108.7003854 \\
UCCSD&	-75.7286759	&-112.4344259&	-75.4220734	&-149.1160634&	-108.6982094\\
UCCSD[T]&	-75.7287535	&-112.4365040 &	-75.4388158	&-149.1191769&	-108.7000018\\
UCCSD(T)	&-75.7287624	&-112.4366356&	-75.4422217	&-149.1192170	&	-108.7000296\\
UCCSD(T*)	&-75.7287535	&-112.4365384	&-75.4409788	&-149.1191937	&	-108.7000143\\
tUCCSD&	-75.7286780	&-112.4347203	&-75.4228943	&-149.1160736	&-108.6982132\\
tUCCSD[T]&	-75.7287561 &	-112.4443853&	-75.4400836&	-149.1193527&	-108.6997618 \\			
tUCCSD(T)	&-75.7287645	&-112.4456299	&-75.4439410	&-149.1193946	&	-108.6997886\\
tUCCSD(T*)	&-75.7287560	&-112.4447561	&-75.4425394	&-149.1193699	&	-108.6997732\\

 \hline
\end{tabular}
\caption{Molecular energies reported in this work.}
\label{table:energies}
\end{center}
\end{table}

\section{Effect of operator ordering}

\begin{figure*}[ht!]
 \centering
 \includegraphics{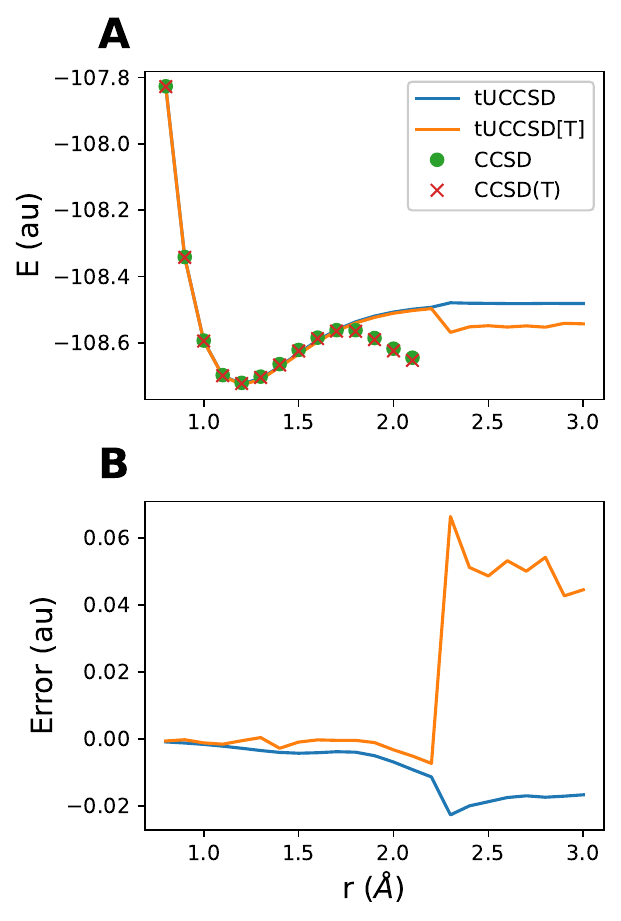}
 \caption{Comparison of the trotterized UCCSD operator and its perturbative, triples correction against their classical CC analogs for the dissociation of the N$_2$ dimer.}
 \label{fig:N2UCC}
\end{figure*}

\begin{figure*}[ht!]
 \centering
 \includegraphics{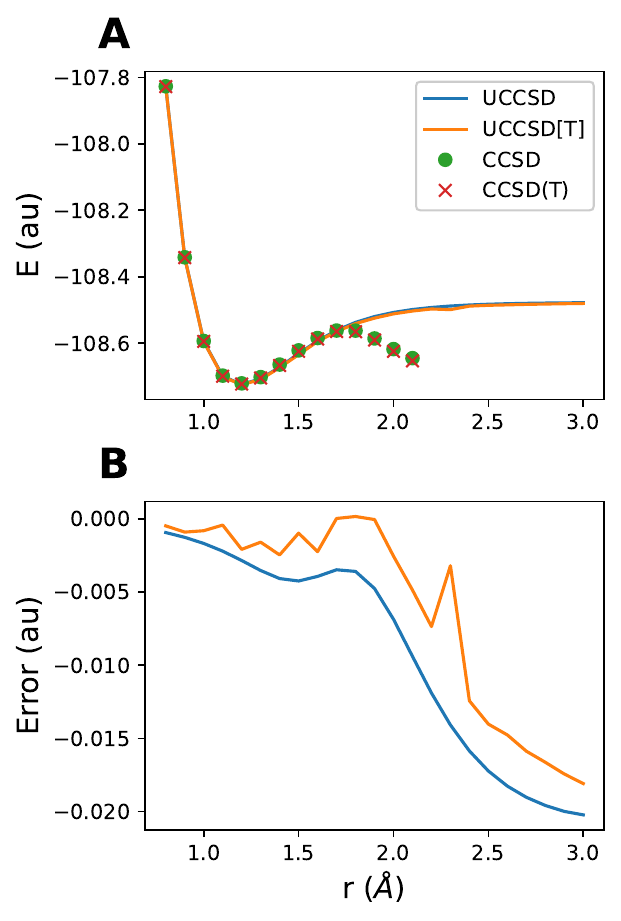}
 \caption{Comparison of the full UCCSD operator and its perturbative, triples correction against their classical CC analogs for the dissociation of the N$_2$ dimer.}
 \label{fig:N2tUCC}
\end{figure*}

Here we provide numerical evidence for the claim that the operator ordering has a sizable effect in the energetics as the N$_2$ departs from the equilibrium region. First, we simply follow the strategy laid out in the main text, we end up Figures \ref{fig:N2UCC} and \ref{fig:N2tUCC}, where we see that tUCCSD and the corresponding [T] do not yield smooth curves, while UCCSD and UCCSD[T] do.

We further analyze this by focusing on energies at $r=2.5$ \AA. For this we report energies with the ordering described in Equation in the main text, its reverse order, and 10 orderings obtained randomly, assigned number 1-10 (these numbers have no particular meaning). The corresponding energies are reported in Table \ref{table:ccsd_shuffle}, on which we only included excitations whose CCSD amplitudes had absolute value above 10$^{-12}$.

\begin{table}[ht!]
\begin{center}
\begin{tabular}{|c| c c c c |} 
 \hline
Ordering	& tUCCSD 	& error (mH)	& tUCCSD[T]	& error (mH) \\ \hline
Original&	-108.4811739	&-18.7954623	&-108.5486183	&48.6489742\\
Reverse	&-108.4852511	&-14.7182743	&-108.4937380	&-6.2313152 \\
1	&-108.4791585	&-20.8108577	&-108.4864910	&-13.4783461\\
2	&-108.4788495	&-21.1198423	&-108.4836705	&-16.2987984\\
3	&-108.481179	&-18.7903034	&-108.4885425	&-11.4268748\\
4	&-108.4731937	&-26.7756670	&-108.4807869	&-19.1824444\\
5	&-108.4738939	&-26.0753895	&-108.4819188	&-18.0505767\\
6	&-108.4705049	&-29.4643874	&-108.4825742	&-17.3951401\\
7	&-108.4816735	&-18.2958133	&-108.4904421	&-9.5272657\\
8	&-108.4749829	&-24.9864422	&-108.4832488	&-16.7205052\\
9	&-108.4827561	&-17.2132415	&-108.4901482	&-9.8211059\\
10	&-108.4764604	&-23.5089757	&-108.4850187	&-14.9506337\\
 \hline
\end{tabular}
\caption{Assessing the impact of operator ordering on the sensitivity of the tUCCSD/[T] result for the N$_2$ molecule at $r=2.5$ \AA. "Original" is the original ordering specified in the manuscript - that depends on the underlying CCSD amplitudes - and "Reverse" is the reverse ordering. Shuffles 1-10 are ordered at random. These results are with respect to an CCSD amplitudes being used as an initial guess to the UCCSD amplitudes. }
\label{table:ccsd_shuffle}
\end{center}
\end{table}

A similar trend persists even if no information is taken from CCSD, which is faulty in this region, as shown in Table \ref{table:shuffle}.

\begin{table}[ht!]
\begin{center}
\begin{tabular}{|c| c c c c |} 
 \hline
Shuffle	& tUCCSD 	& error	(mH)& tUCCSD[T]	& error (mH)\\ \hline
1	&-108.4762272	&-23.7421647	&-108.4817067	&-18.2626184\\
2	&-108.4800833	&-19.8860206	&-108.4931779	&-6.7914257\\
3	&-108.4842005	&-15.7687809	&-108.4967363	&-3.2330415\\
4	&-108.4757842	&-24.1851546	&-108.4887858	&-11.1834957\\
5	&-108.4770212	&-22.9481237	&-108.4884296	&-11.5396819\\
6	&-108.479732	&-20.2373132	&-108.4915446	&-8.4247227\\
7	&-108.4834032	&-16.5661744	&-108.4951588	&-4.8105400\\
8	&-108.4717318	&-28.2375010	&-108.4832757	&-16.6936299\\
9	&-108.4781734	&-21.7959566	&-108.4825076&	-17.4617319\\
10	&-108.4763993	&-23.5699863	&-108.4917173	&-8.2520276\\
 \hline
\end{tabular}
\caption{Assessing the impact of operator ordering on the sensitivity of the tUCCSD/[T] result for the N$_2$ molecule at $r=2.5$ \AA. Shuffles 1-10 are ordered at random. These results are with respect to no initial guess for the UCCSD tau amplitudes.  }
\label{table:shuffle}
\end{center}
\end{table}

\begin{figure*}[ht!]
 \centering
 \includegraphics{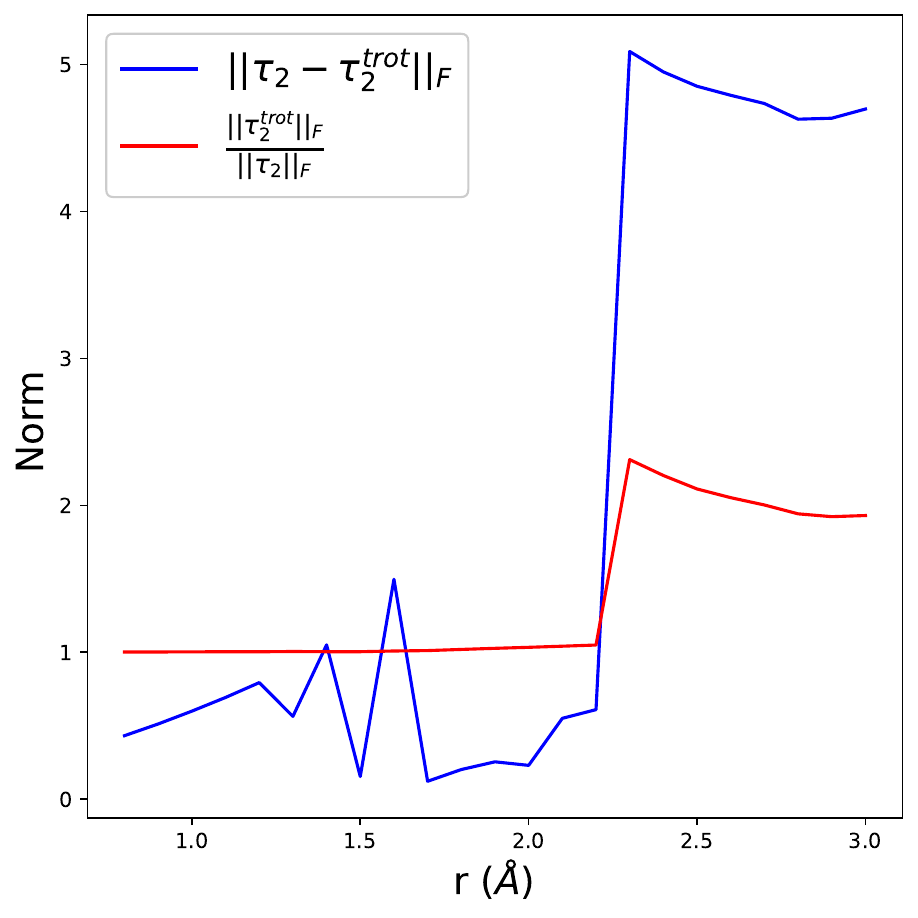}
 \caption{Froebenius norm comparison of the tUCCSD operator and the corresponding full operator along the potential energy surface for N$_2$. $\tau_2$ signifies the full operator, whereas $\tau_2^{trot}$ specifies the Trotterized variant.}
 \label{fig:tau2norm}
\end{figure*}

When we use the ordering shown in Equation in the main text, we can perform two simple comparisons involving the Frobenius norm of the amplitudes along the potential energy surfaces, further displaying the effect of the operator ordering in tUCCSD. We plot those in Figure \ref{fig:tau2norm}.

